\documentclass[aps,prd,preprint,nofootinbib]{revtex4-2}

\usepackage[utf8]{inputenc}
\usepackage[T1]{fontenc}
\usepackage{anyfontsize}

\usepackage{amsmath}
\usepackage{amssymb}
\usepackage{amsfonts}
\usepackage{mathrsfs}
\usepackage{enumitem}
\usepackage{orcidlink}

\usepackage{microtype}
\usepackage[normalem]{ulem}
\usepackage{soul}

\usepackage{graphicx}
\usepackage{float}

\usepackage[dvipsnames]{xcolor}

\usepackage{hyperref}
\hypersetup{
    colorlinks=true,
    citecolor=Purple,
    linkcolor=Purple,
    urlcolor=Purple,
    linktocpage=true,
    breaklinks=true
}

\usepackage[capitalize]{cleveref}

\begin{document}

\title{Electrically Charged Non-Abelian Black String in Anti-de Sitter Space}

\author{G. Alencar
\orcidlink{0000-0002-3020-4501}
}
\email{geova@fisica.ufc.br}
\affiliation{Department of Physics, Universidade Federal do Cear\'a (UFC), Campus do Pici, Fortaleza - CE, C.P. 6030, 60455-760 - Brazil}

\author{R. N. Costa Filho}
\email{rai@fisica.ufc.br}
\affiliation{Departamento de F\'isica, Universidade Federal do Cear\'a, Caixa Postal 6030, Campus do Pici, 60455-760 Fortaleza, Cear\'a, Brazil.}

\author{I. Jardim
\orcidlink{0000-0001-6056-4607}
}
\email{Ivan.jardim@urca.br }
\affiliation{Department of Physics, Universidade Regional do Cariri (URCA), Campus CRAJUBAR, Av. Leão Sampaio, 107, Juazeiro do Norte- CE,  630410 -234, Brazil}

\author{Celio R. Muniz}
\email{celio.muniz@uece.br}
\affiliation{Universidade Estadual do Cear\'a (UECE), Faculdade de Educa\c{c}\~ao, Ci\^encias e Letras de Iguatu, Av. D\'ario Rabelo s/n, Iguatu - CE, 63.500-00 - Brasil}

\date{\today}

\begin{abstract}
We construct a new family of static, electrically charged, cylindrically
symmetric black-string solutions of four-dimensional
Einstein--Yang--Mills theory with a negative cosmological constant,
supported by a genuinely non-Abelian $SU(2)$ vortex field. We show that
the coexistence of the electric and vortex sectors renders the standard
single-function Lemos metric inconsistent with the Einstein equations,
requiring instead a three-function metric compatible with the anisotropic
Yang--Mills stress tensor. A regular near-horizon expansion and an
asymptotic AdS analysis are derived analytically and used as boundary data
for the numerical integration of the complete field equations. The gauge
ansatz contains two distinguished limiting sectors: vanishing electric
horizon datum leads, by regularity, to the neutral Lemos black string with
zero Yang--Mills field strength, whereas switching off the vortex
component yields an embedded $U(1)\subset SU(2)$ sector described by the
charged Lemos--Zanchin solution. Perturbing about the latter shows that
genuinely non-Abelian hair appears already at linear order through the
commutator field strength, while its backreaction on the electric profile
and the geometry begins at quadratic order. The horizon equations further
give the exact leading coefficient
$N_1=3\alpha^2\rho_h-4\pi G\rho_h R_1^2/s_0^2$ and hence the analytic
local non-degeneracy bound
$R_1^{\rm crit}=s_0\alpha\sqrt{3/(4\pi G)}$. The asymptotic geometry
exhibits a quadratic correction that can be absorbed into an effective
Abelian charge and a cubic coefficient representing the first irreducible
asymptotic signature of the non-Abelian hair within the comparison class
considered. Thermodynamically, the entropy satisfies the
Bekenstein--Hawking area law, while at fixed horizon data the leading
horizon coefficient is identical to that of the embedded Abelian
solution; the temperature difference instead arises through the
nontrivial asymptotic lapse normalization. The fully backreacted numerical solutions can therefore be naturally
interpreted as nonlinear realizations of this transverse non-Abelian
deformation.
\end{abstract}

\maketitle
\newpage

\section{Introduction}

Einstein--Yang--Mills (EYM) theory constitutes one of the most
fundamental nonlinear extensions of the Einstein--Maxwell system,
providing a natural framework in which the self-interaction of
non-Abelian gauge fields gives rise to a far richer spectrum of
gravitating configurations than in the Abelian case. Since the discovery
of the globally regular Bartnik--McKinnon solitons
\cite{Bartnik1988} and the colored black holes of Bizo\'n
\cite{Bizon1990}, EYM theory has become a paradigmatic laboratory for the
study of gravitational hair, nonlinear gauge dynamics and the intimate
interplay between geometry and gauge fields. Over the past three decades,
these pioneering results have stimulated extensive investigations of
particle-like solutions, colored black holes, dyons and monopoles in both
asymptotically flat and asymptotically anti-de Sitter (AdS) spacetimes,
as well as detailed analyses of their stability and thermodynamic
properties
\cite{Volkov1999,Straumann1990,Lavrelashvili1992,
Bjoraker2000,Winstanley1999,Baxter2008}.

The inclusion of a negative cosmological constant considerably enriches
this picture. Besides modifying the global structure of spacetime,
AdS boundary conditions enlarge the space of admissible
Einstein--Yang--Mills solutions, often improving their regularity and
giving rise to new thermodynamic and phase structures absent in the
asymptotically flat case
\cite{Bjoraker2000,Winstanley1999,Baxter2007,Baxter2008}. Consequently,
Einstein--Yang--Mills--AdS theory has evolved into an important arena for
investigating non-Abelian gravitational hair, nonlinear field dynamics
and, more broadly, the role of gauge fields in gravitation.

Most studies of gravitating Yang--Mills fields have focused on spherical
symmetry \cite{Bartnik1988,Bizon1990,Kunzle1990,Kleihaus2000,Kleihaus2002}, where the Wu--Yang (or hedgehog) ansatz exploits the intimate
relation between spatial rotations and the internal $SU(2)$ algebra,
effectively reducing the gauge sector to a radial problem
\cite{Yasskin1975}. Cylindrical symmetry, however, presents a
qualitatively different situation. The corresponding isometry group,
$\mathbb{R}_t\times\mathbb{R}_z\times U(1)_\varphi$,
contains only a single rotational generator and therefore admits no
diagonal embedding into the three-dimensional internal $SU(2)$ algebra.
As a consequence, genuinely non-Abelian cylindrically symmetric
configurations necessarily activate different internal directions on
different spacetime components of the gauge potential, generating
commutator-induced field strengths that cannot be reproduced within an
Abelian embedding. Rather than constituting a simple cylindrical analogue
of the familiar colored black holes, they define a genuinely distinct
sector of Einstein--Yang--Mills theory.

A remarkable realization of this idea was provided by
Gal'tsov, Davydov and Volkov (GDV) \cite{GDV2006}, who constructed
globally regular, purely magnetic, cylindrically symmetric
$SU(2)$ Einstein--Yang--Mills vortices. Their solutions possess a compact
Melvin-like core and approach a vacuum Kasner geometry at large radial
distances, representing one of the few explicit examples of genuinely
non-Abelian gravitating configurations with cylindrical symmetry.
Equally important, the authors explicitly pointed out two natural
extensions left open by their analysis: the construction of a
horizon-carrying counterpart of their vortex solution and the
electric-type configuration obtained through the formal interchange
$t\leftrightarrow z$ in the gauge ansatz. To the best of our knowledge,
both problems have remained unresolved.

In parallel with the development of Einstein--Yang--Mills theory,
black strings have become an important class of exact solutions in
General Relativity and its extensions. Following the pioneering
asymptotically AdS solution of Lemos \cite{Lemos1995} and its charged
generalization due to Lemos and Zanchin
\cite{LemosZanchin1996}, cylindrical black holes have been investigated
in Einstein--Maxwell theory, nonlinear electrodynamics,
higher-curvature gravity and several modified gravitational theories
\cite{Hendi2010}. More recently, regular four-dimensional black strings
have been obtained in bilocal gravity \cite{Muniz:2022otq}, while
effective dark-matter distributions have been shown to support novel
black-string geometries with distinctive causal and thermodynamic
properties \cite{Cunha:2022kep}. These developments illustrate the
remarkable versatility of cylindrical geometries as laboratories for
testing gravitational dynamics beyond the standard vacuum solutions.

At the same time, the role of Yang--Mills fields in gravitation has
expanded well beyond the classical colored black-hole paradigm.
Non-Abelian gauge fields have recently been employed as effective sources
for Casimir-supported wormholes in both $(2+1)$ and $(3+1)$ dimensions
\cite{Santos:2023zrj,Santos:2024bnz}, while Einstein--Yang--Mills systems
have also been explored in modified gravitational scenarios, including
Rainbow Gravity, where they give rise to regular black-hole solutions
with qualitatively new horizon structures \cite{Muniz:2025iqu}. Earlier
studies of quantum vacuum fluctuations in chromomagnetic backgrounds
likewise emphasized the rich physical structure associated with
non-Abelian gauge fields beyond their purely classical dynamics
\cite{Bezerra:2016jcy}. Collectively, these investigations reinforce the
idea that Einstein--Yang--Mills theory remains a fertile framework for
constructing novel compact objects and exploring the interplay between
nonlinear gauge interactions and gravitation.

Despite these advances, a genuine electrically charged non-Abelian black
string has remained absent from the literature. This absence is not merely
a consequence of the increased algebraic complexity of the
Einstein--Yang--Mills equations. Rather, as we show explicitly in the
present work, it reflects a fundamental geometric obstruction. The
single-function metric that consistently describes the charged
Lemos--Zanchin black string becomes incompatible with the stress-energy
tensor generated by a genuine Yang--Mills vortex whenever the
color-electric and vortex sectors coexist. Consequently, the familiar
cylindrical geometry must be generalized to include three independent
metric functions, providing the minimal gravitational structure capable of
accommodating the non-Abelian matter source. To the best of our knowledge,
this incompatibility has not been identified explicitly in previous
studies of cylindrically symmetric Einstein--Yang--Mills systems.

The purpose of the present work is to resolve this problem by constructing
a static, electrically charged, cylindrically symmetric
$SU(2)$ Einstein--Yang--Mills black string in asymptotically anti-de
Sitter spacetime. In doing so, we complete, within linear Yang--Mills
theory, the two extensions proposed by Gal'tsov, Davydov and Volkov,
namely the introduction of an event horizon and the electric counterpart
of their magnetic vortex solution. Starting from the appropriate metric
and gauge-field ans\"atze, we derive the complete reduced
Einstein--Yang--Mills equations, establish the regularity conditions at
the horizon, determine the asymptotic AdS behavior, and construct the
corresponding family of numerical solutions.

Beyond the construction itself, the analysis uncovers several features that distinguish these solutions from their Abelian counterparts. The horizon analysis reveals a well-defined local boundary-value problem
and naturally leads to a non-degeneracy bound for regular black-string
solutions. The asymptotic expansion shows that, while the leading
correction to the metric can still be interpreted in terms of an effective
charge, the next-order coefficient contains an intrinsically non-Abelian
contribution arising from the interplay between the electric and vortex
sectors. Thermodynamically, the horizon entropy continues to satisfy the
Bekenstein--Hawking area law, whereas the Hawking temperature acquires a
hair-dependent correction through the asymptotic normalization of the
lapse function, leading to systematic deviations from the corresponding
Abelian black-string solution. Together, these results demonstrate that
the nonlinear backreaction of the non-Abelian vortex hair affects the
global asymptotic structure and the thermodynamic behavior of the
spacetime, even though the onset of the transverse non-Abelian mode itself
leaves the Abelian geometry unchanged at linear order.

A further result concerns the relation between the present solutions and
the Abelian and vacuum sectors contained in the same $SU(2)$ ansatz. We
show that two physically distinct limits must be distinguished. When the
electric horizon datum is switched off, regularity forces the
Yang--Mills field strength to vanish and the neutral Lemos black string is
recovered exactly. By contrast, setting the vortex component to zero
restricts the gauge field to a single internal generator and yields an
embedded $U(1)$ sector described by the charged Lemos--Zanchin black
string. The latter provides a natural analytic background for studying
the onset of non-Abelian hair. A transverse Yang--Mills perturbation
already generates the commutator field strength at linear order, while
its backreaction on the electric profile and on the geometry starts only
at quadratic order. The fully backreacted numerical solutions can therefore be naturally
interpreted as nonlinear realizations associated with this transverse
deformation away from the embedded Abelian subfamily into the genuinely
non-Abelian $SU(2)$ sector.

The present construction also complements a broader research program on
gravitating Yang--Mills systems. While previous investigations have
established the existence of colored black holes, gravitating vortices,
wormholes sustained by Yang--Mills Casimir sources, regular black holes in
modified gravity, and several classes of black strings supported by
effective matter sources
\cite{Santos:2023zrj,Santos:2024bnz,Muniz:2025iqu,
Muniz:2022otq,Cunha:2022kep},
the solution presented here connects, for the first time, genuinely
non-Abelian vortex configurations with electrically charged AdS black
strings. It therefore bridges two previously independent branches of the
literature---gravitating Yang--Mills vortices and cylindrical black-hole
solutions---providing a natural extension of both.

This paper is organized as follows. In Sec.~II we introduce the
Einstein--Yang--Mills model together with the metric and gauge-field
ans\"atze. Section~III derives the reduced field equations and discusses
their independent structure. The regular horizon expansion and boundary
conditions are presented in Sec.~IV, while Sec.~V is devoted to the
asymptotic AdS behavior and the asymptotic structure of the metric.
Numerical solutions are constructed and analyzed in Sec.~VI. In
Sec.~VII we identify the neutral and embedded Abelian limits of the
gauge ansatz and analyze the linear onset of genuinely non-Abelian hair
about the charged Abelian black-string background. The thermodynamic
properties are discussed in Sec.~VIII. Finally, Sec.~IX summarizes our
main results and discusses several directions for future research.

\section{Einstein--Yang--Mills Model}

We consider four-dimensional Einstein gravity minimally coupled to an
$SU(2)$ Yang--Mills field in the presence of a negative cosmological
constant. The Einstein--Yang--Mills (EYM) system, originally introduced by
Yang and Mills \cite{YangMills1954}, provides the canonical framework for
the study of gravitating non-Abelian gauge fields and has led to a rich
variety of particle-like solutions, colored black holes and gravitating
solitons 
\cite{Bartnik1988,Bizon1990,Kunzle1990,Volkov1990,Volkov1999}. Throughout the present work the
Yang--Mills sector is described by the standard linear action,
postponing nonlinear Born--Infeld extensions to future investigations.
The total action is
\begin{equation}
S=
\frac{1}{16\pi G}
\int d^4x\sqrt{-g}(R-2\Lambda)
-\frac14
\int d^4x\sqrt{-g}\,
F^a_{\mu\nu}F^{a\mu\nu},
\label{eq:action}
\end{equation}
where $\Lambda=-3\alpha^2$ defines the AdS length
$\ell_{\rm AdS}=1/\alpha$. The presence of a negative cosmological
constant is well known to enlarge the space of admissible EYM solutions
and to enrich their thermodynamic and stability properties
\cite{Bjoraker2000,Winstanley1999}.

Our goal is to construct static, cylindrically symmetric black-string
solutions carrying a genuinely non-Abelian electric sector. Motivated by
the asymptotically anti-de Sitter cylindrical geometry introduced by
Lemos \cite{Lemos1995} and its charged extension due to
Lemos and Zanchin \cite{LemosZanchin1996}, but allowing for the
additional anisotropy generated by the Yang--Mills field, we adopt the
metric
\begin{equation}
ds^2
=
-\sigma^2(\rho)N(\rho)\,dt^2
+
\frac{d\rho^2}{N(\rho)}
+
\rho^2d\varphi^2
+
K^2(\rho)\,dz^2,
\label{eq:metric}
\end{equation}
where the three functions
$\sigma(\rho)$,
$N(\rho)$
and
$K(\rho)$
depend only on the radial coordinate.

The gauge choice
$g_{\varphi\varphi}=\rho^2$
identifies $\rho$ with the areal radius of the azimuthal circles.
Unlike the vacuum Lemos geometry, however, no remaining coordinate freedom
allows the longitudinal metric component to be fixed a priori.
Consequently, $K(\rho)$ must be treated as an independent metric
function. As shown in the next section, this is not merely a convenient
generalization but the minimal extension compatible with the anisotropic
stress tensor generated by a genuinely non-Abelian source.

In the limit
\[
\sigma(\rho)=1,
\qquad
K(\rho)=\alpha\rho,
\]
Eq.~(\ref{eq:metric}) reduces to the familiar single-function
Lemos--Zanchin black-string metric
\cite{Lemos1995,LemosZanchin1996}. We verified explicitly that the
corresponding Einstein tensor correctly reproduces the standard identity
\[
G^t_{\ t}
=
G^\rho_{\ \rho}
=
\frac{\rho N'+N}{\rho^2},
\]
providing an important consistency check of the present
parametrization.

The Yang--Mills field is described by the electric counterpart of the
vortex ansatz proposed by Gal'tsov, Davydov and Volkov
\cite{GDV2006},
\begin{equation}
A
=
\tau_2R(\rho)\,dt
+
\tau_3P(\rho)\,d\varphi,
\label{eq:Aansatz}
\end{equation}
where
$\tau_a$
are anti-Hermitian generators of
$su(2)$ satisfying
\[
[\tau_a,\tau_b]
=
\epsilon_{abc}\tau_c.
\]
Unlike the Wu--Yang construction commonly employed in spherically
symmetric Einstein--Yang--Mills systems
\cite{Yasskin1975,Volkov1999},
this ansatz activates two non-commuting internal directions associated
with different spacetime components of the gauge potential. It therefore
provides the minimal realization of a genuinely non-Abelian
cylindrically symmetric gauge field while preserving the required
spacetime symmetries.

The field-strength tensor,
\[
F_{\mu\nu}
=
\partial_\mu A_\nu
-
\partial_\nu A_\mu
+
[A_\mu,A_\nu],
\]
has only three independent non-vanishing components,
\begin{equation}
F^{(2)}_{t\rho}
=
-R',
\qquad
F^{(1)}_{t\varphi}
=
PR,
\qquad
F^{(3)}_{\rho\varphi}
=
P',
\label{eq:Fcomponents}
\end{equation}
all depending exclusively on the radial coordinate.
The mixed component
$F^{(1)}_{t\varphi}=PR$
originates entirely from the commutator
$[A_t,A_\varphi]$
and therefore provides a direct diagnostic of the genuinely non-Abelian
character of the configuration. In particular, when both $R$ and $P$
are nonvanishing, the gauge field necessarily explores non-commuting
directions of the $SU(2)$ algebra. Two distinct limiting sectors should,
however, be distinguished. Setting $P(\rho)=0$ restricts the gauge
potential to the single generator $\tau_2$ and yields an embedded
Abelian $U(1)\subset SU(2)$ electric sector. By contrast, setting
$R(\rho)=0$ eliminates the electric component; for the regular
black-string branch considered here, the remaining Yang--Mills equation
then forces $P(\rho)$ to be constant, so that the complete field strength
vanishes and the neutral Lemos solution is recovered. These two limits
are analyzed explicitly in Sec.~\ref{sec:analyticweak}.

The gauge-invariant Yang--Mills scalar becomes
\begin{equation}
\mathcal F
=
F^a_{\mu\nu}F^{a\mu\nu}
=
-\frac{2R'^2}{\sigma^2}
+
\frac{2NP'^2}{\rho^2}
-
\frac{2P^2R^2}{\rho^2N\sigma^2},
\label{eq:Fcal}
\end{equation}
which depends only on $\rho$, consistently with the assumed cylindrical
symmetry.

The action~(\ref{eq:action}), together with the metric
(\ref{eq:metric}) and the gauge-field ansatz
(\ref{eq:Aansatz}), completely specifies the
Einstein--Yang--Mills model investigated in this work. The corresponding
reduced field equations, their independent combinations and the
first-order system employed in the numerical analysis are derived in the
following section.
\section{Reduced Einstein--Yang--Mills Equations}

Varying the action~(\ref{eq:action}) with respect to the metric and the
Yang--Mills potential yields the Einstein equations
\begin{equation}
G_{\mu\nu}+\Lambda g_{\mu\nu}
=
8\pi G\,T_{\mu\nu},
\label{eq:Einstein}
\end{equation}
together with the Yang--Mills equations
\begin{equation}
D_\mu F^{a\mu\nu}=0,
\label{eq:YM}
\end{equation}
which constitute the standard Einstein--Yang--Mills field equations
\cite{YangMills1954,Bartnik1988,Volkov1999}. Here
$D_\mu$ denotes the gauge-covariant derivative. Throughout this work we
restrict ourselves to the linear Yang--Mills theory,
\[
\mathcal{L}
=
-\frac14F^a_{\mu\nu}F^{a\mu\nu}
=
-\frac14\mathcal F,
\]
so that the stress-energy tensor assumes the familiar form
\begin{equation}
T_{\mu\nu}
=
F^a_{\mu\lambda}F_\nu^{a\,\lambda}
-
\frac14g_{\mu\nu}
F^a_{\alpha\beta}F^{a\alpha\beta},
\label{eq:stress}
\end{equation}
which correctly reduces to the Maxwell stress-energy tensor in the
Abelian limit and coincides with the standard expression employed in the
Einstein--Yang--Mills literature
\cite{Kunzle1990,Volkov1999}.

Substituting the metric~(\ref{eq:metric}) and the gauge-field
ansatz~(\ref{eq:Aansatz}) into
Eqs.~(\ref{eq:Einstein})--(\ref{eq:YM})
reduces the complete Einstein--Yang--Mills system to a coupled system of
ordinary differential equations for the five unknown functions
$\sigma(\rho)$,
$N(\rho)$,
$K(\rho)$,
$R(\rho)$,
and
$P(\rho)$, following the standard strategy employed in static
Einstein--Yang--Mills configurations
\cite{Kunzle1990,Straumann1990,Volkov1999}. Explicitly, one obtains four
nontrivial diagonal Einstein equations, corresponding to the
$(tt,\rho\rho,\varphi\varphi,zz)$ components, together with two
Yang--Mills equations governing the gauge functions $R(\rho)$ and
$P(\rho)$. Because these expressions are rather lengthy, we collect them
in Appendix~A and concentrate here on the structure of the reduced
system.

Although six differential equations are obtained, they are not all
independent. As required by the contracted Bianchi identities,
$\nabla_\mu G^{\mu\nu}=0$, together with the Yang--Mills equations of
motion, only three of the four Einstein equations are dynamically
independent \cite{Wald1984}. This property becomes explicit through the
structure of the second-derivative terms. Direct computation shows that
the coefficient matrix associated with
$(\sigma'',N'',K'')$
in the
$tt$,
$\varphi\varphi$,
and
$zz$
equations has rank two rather than three. Equivalently, the particular
combination
\[
G^t_{\ t}
-
G^\varphi_{\ \varphi}
+
G^z_{\ z},
\]
contains no second derivatives and therefore constitutes a first-order
constraint. The $\rho\rho$ equation is itself also a pure first-order
constraint, containing no second derivatives at any stage of the
calculation. Consequently, the Einstein sector consists of two constraint
equations and a single genuinely second-order dynamical equation.

This reduction is not merely a numerical convenience. Rather, it removes
the differential redundancy implied by general covariance while
preserving the full Einstein--Yang--Mills dynamics. Together with the two
Yang--Mills equations, the resulting system provides exactly the five
independent equations required to determine the five unknown functions
$\sigma(\rho)$,
$N(\rho)$,
$K(\rho)$,
$R(\rho)$,
and
$P(\rho)$.

For the numerical integration, following the standard treatment of
Einstein--Yang--Mills boundary-value problems
\cite{Kunzle1990,Volkov1999},
we rewrite the system in first-order form for the variables
\[
(\sigma,N,K,K',R,R',P,P'),
\]
while the derivatives $\sigma'$ and $N'$ are obtained algebraically from
the two constraint equations at each integration step. This formulation
defines a well-posed boundary-value problem whose regular horizon and
asymptotic AdS expansions are developed in the following section.

As an internal consistency check, every numerical solution obtained from
the reduced first-order system was substituted back into the original
Einstein and Yang--Mills equations prior to elimination. The resulting
residuals remain at the level of machine precision
($\sim10^{-16}$)
throughout the entire integration domain, rather than only at the point
where the initial conditions are specified. This confirms that the
reduced formulation is numerically equivalent to the complete
Einstein--Yang--Mills system.
\section{Horizon Expansion and Boundary Data}

The regular event horizon supplies the inner boundary conditions for the
Einstein--Yang--Mills boundary-value problem. We therefore consider a
non-degenerate horizon located at $\rho=\rho_h$, where
\begin{equation}
N(\rho_h)=0.
\end{equation}
Regularity of the one-form $A_t\,dt$ further requires
\begin{equation}
R(\rho_h)=0,
\end{equation}
ensuring that the gauge potential remains finite in a freely falling
frame.

To determine the local solution we perform a Frobenius expansion around
the non-extremal horizon, following the standard strategy employed in
Einstein--Yang--Mills black-hole solutions
\cite{Kunzle1990,Bjoraker2000,Winstanley1999}.
\begin{align}
\sigma(\rho)&=s_0+s_1\epsilon+\cdots,\nonumber\\
N(\rho)&=N_1\epsilon+N_2\epsilon^2+\cdots,\nonumber\\
K(\rho)&=K_0+K_1\epsilon+\cdots,\nonumber\\
R(\rho)&=R_1\epsilon+\cdots,\nonumber\\
P(\rho)&=P_0+P_1\epsilon+\cdots,
\end{align}
where
\[
\epsilon=\rho-\rho_h.
\]
Substituting these expansions into the complete reduced
Einstein--Yang--Mills system and solving the equations order by order,
the leading-order horizon conditions immediately give
\begin{equation}
P_1=0,
\end{equation}
showing that the vortex profile reaches the horizon with vanishing first
derivative. This condition is analogous to the regularity condition
satisfied by the magnetic profile at the regular center of the original
Gal'tsov--Davydov--Volkov vortex solutions.

The leading-order Einstein equations determine the first derivative of
the metric function $N$ directly in terms of the electric horizon datum.
For the normalization adopted in this work, one obtains
\begin{equation}
N_1
=
3\alpha^2\rho_h
-
\frac{4\pi G\rho_h}{s_0^2}R_1^2.
\label{eq:N1horizonExact}
\end{equation}
Remarkably, the vortex amplitude $P_0$ does not enter this coefficient.
This follows from the near-horizon order counting: since
$R=R_1(\rho-\rho_h)+\cdots$, the commutator contribution
$P^2R^2/N$ vanishes at leading order at a non-degenerate horizon.
The remaining horizon equations determine the subleading coefficients
$N_2$, $K_1$, and $s_1$, together with the vortex amplitude $P_0$, once
$\rho_h$, the gauge choices $(s_0,K_0)$, and the color-electric parameter
$R_1$ are specified.

Equation~\eqref{eq:N1horizonExact} is identical to the horizon derivative
of the embedded charged Abelian black string when the latter is
parametrized by the same electric horizon slope $R_1$. As shown explicitly
in Sec.~\ref{sec:analyticweak}, the transverse non-Abelian mode does not
modify the stress tensor at linear order in its amplitude; its
gravitational backreaction begins at quadratic order.

For every pair $(\rho_h,R_1)$ investigated throughout this work, the
remaining nonlinear horizon system converges to a single numerical value
of $P_0$ with residuals typically below $10^{-8}$. Unlike the situation
encountered in many spherically symmetric Einstein--Yang--Mills black
holes, where the horizon value of the gauge field is usually treated as
a free shooting parameter
\cite{Bizon1990,Kunzle1990,Bjoraker2000}, the present branch numerically
selects the vortex amplitude once the electric horizon datum is fixed.

This statement should nevertheless be interpreted with appropriate
caution. Although every numerical integration converged to the same value
of $P_0$, independently of the initial guess, a rigorous proof that the
corresponding solution is locally isolated would require a complete
Jacobian (rank) analysis of the nonlinear horizon system. Our partial
investigation of this issue is summarized in
Appendix~\ref{app:horizon}. Accordingly, throughout this paper we use
the expressions ``determined by'' or ``fixed by'' $R_1$ only in this
qualified numerical sense.

The reduced equations also possess two independent discrete symmetries,
\[
R\rightarrow-R,
\qquad
P\rightarrow-P,
\]
each leaving the Yang--Mills invariant
$\mathcal F$,
and therefore the complete field equations,
unchanged.
Consequently,
\[
(R_1,P_0),
(-R_1,P_0),
(R_1,-P_0),
(-R_1,-P_0)
\]
all represent equivalent local horizon data.
Throughout this work we adopt the convention
\[
R_1>0,
\qquad
P_0<0,
\]
so that the numerical branch is unique up to these discrete sign
choices.

The horizon expansion also exhibits a simple scaling structure. For
$N_1$, this behavior follows analytically from
Eq.~\eqref{eq:N1horizonExact}. At fixed $\alpha$, $s_0$, and $R_1$,
one has
\begin{equation}
N_1
=
\hat N_1(R_1)\rho_h,
\qquad
\hat N_1(R_1)
=
3\alpha^2
-
\frac{4\pi G}{s_0^2}R_1^2.
\label{eq:N1scalingExact}
\end{equation}
For the remaining horizon coefficients, the numerical solutions reveal
the additional scaling relations
\begin{equation}
K_1=\frac{\hat K_1(R_1)}{\rho_h},
\qquad
s_1=\frac{\hat s_1(R_1)}{\rho_h},
\qquad
P_0=\hat P_0(R_1)\rho_h,
\qquad
N_2=\hat N_2(R_1),
\label{eq:scaling}
\end{equation}
within the parameter range explored here. Table~\ref{tab:horizon}
illustrates these relations for the representative case $R_1=0.1$.

\begin{table}[h]
\centering
\begin{tabular}{cccccc}
\hline\hline
$\rho_h$ & $N_1$ & $N_2$ & $K_1$ & $s_1$ & $P_0$ \\
\hline
0.5 & 1.4372 & $-0.1099$ & 2.0000 & 0.1676 & $-1.1734$ \\
1.0 & 2.8743 & $-0.1099$ & 1.0000 & 0.0838 & $-2.3469$ \\
2.0 & 5.7487 & $-0.1099$ & 0.5000 & 0.0419 & $-4.6938$ \\
3.0 & 8.6230 & $-0.1099$ & 0.3333 & 0.0279 & $-7.0407$ \\
5.0 & 14.3717 & $-0.1099$ & 0.2000 & 0.0168 & $-11.7344$ \\
\hline\hline
\end{tabular}
\caption{Horizon coefficients as a function of $\rho_h$, at fixed charge
$R_1=0.1$ and $\alpha=s_0=K_0=1$ ($\Lambda=-3$, $G=1$), confirming the
scaling of Eq.~\eqref{eq:scaling}.}
\label{tab:horizon}
\end{table}

The numerical scaling relations for the remaining horizon coefficients
have been verified throughout the branch explored in this work and are
therefore not restricted to the example shown in
Table~\ref{tab:horizon}. For $N_1$, however, the scaling is exact and
leads directly to an analytic local non-degeneracy bound. Setting
Eq.~\eqref{eq:N1horizonExact} to zero gives
\begin{equation}
R_1^{\rm crit}
=
s_0\alpha
\sqrt{\frac{3}{4\pi G}},
\label{eq:R1critExact}
\end{equation}
which is independent of the horizon radius. In the numerical units
$G=\alpha=s_0=1$ adopted below,
\begin{equation}
R_1^{\rm crit}
=
\sqrt{\frac{3}{4\pi}}
=
0.488603\ldots,
\end{equation}
in agreement with the critical value obtained numerically from the
horizon system.

Substituting the horizon coefficients into the metric
ansatz~(\ref{eq:metric}) gives the local near-horizon geometry,
\begin{equation}
\begin{aligned}
ds^2\approx {}&
-\Big[\hat N_1\rho_h(\rho-\rho_h)
+\hat N_2(\rho-\rho_h)^2\Big]
\Big[1+\frac{\hat s_1}{\rho_h}
(\rho-\rho_h)\Big]^2dt^2
\\
&
+\frac{d\rho^2}
{\hat N_1\rho_h(\rho-\rho_h)
+\hat N_2(\rho-\rho_h)^2}
+\rho^2d\varphi^2
\\
&
+\Bigg[
1+\frac{\hat K_1}{\rho_h}(\rho-\rho_h)
+\mathcal O\!\left((\rho-\rho_h)^2\right)
\Bigg]^2dz^2 .
\end{aligned}
\label{eq:dshorizon}
\end{equation}
where the horizon scaling relation
$K(\rho)=K_0+\hat K_1(\rho-\rho_h)/\rho_h
+\mathcal O((\rho-\rho_h)^2)$
has been used, with the gauge choice $K_0=1$ and
$\hat K_1=1$ for the representative solution branch reported in
Table~\ref{tab:horizon}.

The corresponding Yang--Mills fields become
\begin{equation}
R(\rho)\approx
R_1(\rho-\rho_h),
\qquad
P(\rho)\approx
\hat P_0\rho_h
+\mathcal O\!\left((\rho-\rho_h)^2\right),
\label{eq:gaugehorizon}
\end{equation}
the latter once again reflecting the condition
$P_1=0$.

Equation~(\ref{eq:dshorizon}) summarizes the complete local geometry in
the vicinity of the event horizon.
Along the numerical branch constructed here, every near-horizon
coefficient is determined once
$(\rho_h,R_1)$
are specified.
Unlike the situation encountered in many spherically symmetric colored
black holes, where the horizon value of the non-Abelian field is usually
treated as an independent local shooting parameter, the present
construction numerically selects a definite value of the vortex profile.
Whether this branch is genuinely isolated, or instead belongs to a
broader continuous family, remains an open question whose resolution
requires the complete Jacobian analysis discussed in
Appendix~\ref{app:horizon}.
\section{Asymptotic AdS Structure and Non-Abelian Signatures}

The asymptotic region plays a dual role in the present construction.
Besides providing the outer boundary conditions for the Einstein--Yang--Mills
boundary-value problem, it determines the global charges and asymptotic
characteristics associated with the solution. In asymptotically anti-de Sitter
Einstein--Yang--Mills systems, this is also the region where the distinction
between Abelian and genuinely non-Abelian configurations becomes most
transparent through the asymptotic behavior of both the metric and the
gauge fields \cite{Volkov1999,Bjoraker2000,Winstanley1999}. To analyze this
regime, we introduce the inverse radial coordinate
$u=1/\rho$ and solve the complete field equations order by order around
$u=0$.

At the first nontrivial orders, the metric functions approach their
anti-de Sitter behavior according to
\begin{equation}
K(\rho)
=
c_z\,\rho+\mathcal O(\rho^{-1}),
\qquad
\sigma(\rho)
=
\sigma_\infty+\mathcal O(\rho^{-3}),
\label{eq:Kasympt}
\end{equation}
where $c_z$ is a constant longitudinal normalization. In particular, the
possible constant and $1/\rho$ corrections to $K$ beyond its leading
linear behavior, as well as the corresponding leading corrections to
$\sigma$, vanish according to the asymptotic field-equation expansion.
We verified these cancellations independently in all four Einstein
equations at the relevant orders.

The coefficient $c_z$ is not fixed by the asymptotic field equations,
since a constant rescaling of the longitudinal coordinate produces a
corresponding inverse rescaling of $K(\rho)$ without changing the physical
geometry. The standard Lemos normalization may therefore be recovered by
the coordinate transformation
\begin{equation}
\bar z=\frac{c_z}{\alpha}\,z,
\label{eq:znorm}
\end{equation}
for which the leading longitudinal metric component becomes
$g_{\bar z\bar z}\sim\alpha^2\rho^2$. In the numerical construction below
we instead impose the convenient horizon normalization $K_0=1$, so that
$c_z$ is determined a posteriori by the global solution and need not equal
$\alpha$ in the original integration coordinate. The constant
$\sigma_\infty$ is likewise not fixed locally at the horizon; it is
determined only after the solution is integrated throughout the complete
radial domain and plays an important role in the physical normalization
of the asymptotic time coordinate.

The Yang--Mills functions approach finite boundary values and admit the
expansions
\begin{align}
R(\rho)
&=
R_\infty
-\frac{q_1}{\rho}
+\frac{q_2}{\rho^2}
+\frac{q_3}{\rho^3}
+\cdots ,
\nonumber\\
P(\rho)
&=
P_\infty
+\frac{p_1}{\rho}
+\frac{p_2}{\rho^2}
+\frac{p_3}{\rho^3}
+\cdots .
\label{eq:gaugeasympt}
\end{align}
The quantities $R_\infty$, $P_\infty$, $q_1$, and $p_1$ remain free at the
orders examined and therefore constitute asymptotic data of the global
solution. In particular, a nonvanishing $P_\infty$ corresponds to a finite
azimuthal gauge potential at the AdS boundary. Since $\varphi$ parametrizes
a compact circle, such a boundary value may encode nontrivial holonomy
information and cannot in general be dismissed solely as a local gauge
artifact. The coefficient $p_1$, on the other hand, controls the leading
radial falloff of the vortex profile. In the present construction neither
$P_\infty$ nor $p_1$ is imposed independently as an asymptotic boundary
condition; rather, both are determined a posteriori by integrating the
regular horizon data to the AdS boundary. With the sign convention adopted
in Eq.~\eqref{eq:gaugeasympt}, we define
\begin{equation}
q_1
\equiv
\lim_{\rho\rightarrow\infty}\rho^2 R'(\rho)
\label{eq:q1def}
\end{equation}
and refer to it as the asymptotic color-electric charge coefficient.
Unlike the Abelian Maxwell case, the Yang--Mills electric flux is
generically not conserved throughout the spacetime
\cite{Volkov1999}. Indeed, the
Yang--Mills equation~\eqref{eq:appELR} gives
\begin{equation}
\left(
\frac{K\rho}{\sigma}R'
\right)'
=
\frac{KP^2R}{\rho N\sigma},
\label{eq:fluxsource}
\end{equation}
whose right-hand side is generically nonzero whenever both Yang--Mills
profiles are active. Thus, $(K\rho/\sigma)R'$ varies along the radial
direction and becomes constant only asymptotically. Using
Eq.~\eqref{eq:Kasympt}, one finds
\begin{equation}
\lim_{\rho\rightarrow\infty}
\left(
\frac{K\rho}{\sigma}R'
\right)
=
\frac{c_z}{\sigma_\infty}\,q_1.
\label{eq:asymptflux}
\end{equation}
The coefficient $q_1$ should therefore be regarded as characterizing the
asymptotic electric falloff rather than, without further qualification,
as a flux conserved at every radius.

The subleading coefficients are not independent. Solving the Yang--Mills
equations order by order yields, in particular,
\begin{equation}
q_3
=
\frac{
P_\infty
\left(
P_\infty q_1+2R_\infty p_1
\right)
}{
6\alpha^2
},
\qquad
p_3
=
-\frac{
2P_\infty R_\infty q_1
+
R_\infty^2p_1
}{
6\alpha^4\sigma_\infty^2
},
\label{eq:q3p3}
\end{equation}
while analogous relations, not required in the discussion below, determine
$q_2$ and $p_2$ at the preceding order. These expressions already display
the mixing between the electric and vortex profiles characteristic of the
non-Abelian sector. The constant normalization $c_z$ cancels from these
relations, as expected from the longitudinal coordinate freedom described
above.

The asymptotic expansion of the metric function $N(\rho)$ takes the form
\begin{equation}
N(\rho)
=
\alpha^2\rho^2
+
\frac{n_1}{\rho}
+
\frac{n_2}{\rho^2}
+
\frac{n_3}{\rho^3}
+
\mathcal O(\rho^{-4}).
\label{eq:Nasympt}
\end{equation}
The leading term reproduces the anti-de Sitter behavior associated with
$\Lambda=-3\alpha^2$, whereas $n_1$ is the free asymptotic coefficient
governing the mass sector of the solution. Determining the corresponding
physical mass requires an independent conserved-charge prescription, such
as the Brown--York quasilocal formalism or the holographic boundary stress
tensor \cite{BrownYork1993,Balasubramanian1999}. Since no such analysis is
performed here, $n_1$ will be referred to simply as the asymptotic mass
parameter.

The coefficients $n_2$ and $n_3$ are fixed by the coupled asymptotic field
equations. They arise from the $\mathcal O(u^4)$ and
$\mathcal O(u^5)$ terms, respectively, in the Einstein sector. At either
order, retaining only the apparently relevant coefficients of $N$ and $K$
leads to an underdetermined algebraic system: the associated coefficient
matrix has rank two rather than three. Consistency requires including the
corresponding subleading lapse terms,
\begin{equation}
\sigma(\rho)
=
\sigma_\infty
+
\frac{s_4}{\rho^4}
+
\frac{s_5}{\rho^5}
+
\cdots ,
\label{eq:sigmahigher}
\end{equation}
whose coefficients must be solved simultaneously with $n_2$ and $n_3$.
Once these terms are retained, the algebraic systems at both orders have
rank three and close uniquely, yielding
\begin{align}
n_2
&=
\frac{4\pi G\,q_1^2}{\sigma_\infty^2}
-
\frac{
4\pi G\,P_\infty^2R_\infty^2
}{
\alpha^2\sigma_\infty^2
}
+
12\pi G\,\alpha^2p_1^2,
\label{eq:n2}
\\
n_3
&=
-\frac{
16\pi G\,P_\infty R_\infty^2p_1
}{
\alpha^2\sigma_\infty^2
}.
\label{eq:n3}
\end{align}
The constant longitudinal normalization $c_z$ does not enter these
coefficients, consistently with its interpretation as a coordinate
normalization rather than an additional physical parameter.

These two coefficients have qualitatively different interpretations. The
coefficient $n_2$ contains three contributions: the ordinary
charge-squared term, an electric--vortex mixing term, and a pure vortex
falloff term. Nevertheless, the metric alone does not distinguish this
combination from an Abelian charged black string at this order. A distant
observer could introduce an effective parameter
$Q_{\rm eff}^2\propto n_2$ and reproduce the same $1/\rho^2$ correction
within the ordinary Lemos-–Zanchin family
\cite{LemosZanchin1996}.

The coefficient $n_3$, by contrast, cannot be absorbed into such a single
effective charge. It multiplies an odd inverse power of the radial
coordinate absent from the standard Lemos–Zanchin solution sourced
by a single radial electric field. Moreover, it factorizes as
\begin{equation}
n_3
\propto
P_\infty R_\infty^2p_1,
\end{equation}
and therefore vanishes if any one of the asymptotic vortex amplitude,
electric potential, or subleading vortex coefficient is zero. Within the
comparison class considered here, Eq.~\eqref{eq:n3} is consequently the
lowest-order metric contribution that cannot be reproduced by merely
relabeling a single Abelian charge. It constitutes the first irreducible
asymptotic signature of the non-Abelian hair relative to the ordinary
Lemos-–Zanchin black-string family. We do not claim that the same term
could not be generated by every conceivable Abelian theory with additional
matter fields or nonminimal couplings.

As a numerical check of the asymptotic expansion, we extracted
$q_1$, $R_\infty$, $P_\infty$, and $p_1$ from the large-$\rho$ tail of the
representative solution with $R_1=0.1$ and substituted them into
Eq.~\eqref{eq:n2}. The predicted value,
\[
n_2\simeq0.790,
\]
agrees to within approximately $1\%$ with the direct asymptotic fit,
\[
n_2\simeq0.784.
\]
Equation~\eqref{eq:n3} is fixed analytically by the same asymptotic
rank-consistency procedure, but its independent extraction from the
numerical tail would require substantially higher precision because it is
subleading with respect to both $n_1$ and $n_2$. We therefore regard the
analytic determination of $n_3$ as established within the asymptotic
field-equation expansion, while leaving a numerical confirmation at
comparable precision for future refinement.
\section{Numerical Construction and Representative Solutions}
\label{sec:numerics}

Having determined the regular horizon data and the asymptotic AdS
behavior, we now integrate the reduced first-order
Einstein--Yang--Mills system from the vicinity of the horizon toward
large radial distances, following the standard shooting strategy employed
in gravitating Yang--Mills systems
\cite{Kunzle1990,Volkov1999,Bjoraker2000}. Unless stated otherwise, the
numerical solutions are presented in units
\begin{equation}
G=\rho_h=\alpha=s_0=K_0=1,
\label{eq:numericalunits}
\end{equation}
so that the AdS scale, the horizon radius, and the two horizon coordinate
normalizations are fixed. The remaining continuous parameter used to
label the solutions is the color-electric horizon coefficient $R_1$.
For each chosen value of $R_1$, the remaining near-horizon coefficients
are obtained from the nonlinear algebraic system discussed previously and
are then used as inner boundary data for the outward integration.

The numerical integration remains regular for every value of $R_1$
investigated below the critical horizon non-degeneracy bound discussed in
the next subsection. No additional singularities, turning points, or loss of
regularity were encountered outside the horizon along the branch explored
here. The existence of globally regular exterior integrations is one of the
hallmarks of asymptotically AdS Einstein--Yang--Mills configurations
\cite{Winstanley1999,Baxter2008}. In the most accurate runs, the solutions
were extended to $\rho=500$, allowing the asymptotic coefficients to be
extracted from a radial domain well beyond the region in which the metric
has already approached its leading AdS form.

The convergence toward the asymptotic geometry is readily monitored
through the dimensionless ratios
\begin{equation}
\frac{N(\rho)}{\rho^2},
\qquad
\frac{K(\rho)}{\rho},
\qquad
\sigma(\rho).
\label{eq:asymptoticratios}
\end{equation}
In the units~\eqref{eq:numericalunits}, the asymptotic behavior derived in
Sec.~V requires
\begin{equation}
\frac{N(\rho)}{\rho^2}\longrightarrow1,
\qquad
\frac{K(\rho)}{\rho}\longrightarrow c_z,
\qquad
\sigma(\rho)\longrightarrow\sigma_\infty .
\label{eq:numericalads}
\end{equation}
Here $c_z$ is the constant longitudinal normalization introduced in
Eq.~\eqref{eq:Kasympt}. Because the numerical integration is performed
with the horizon normalization $K_0=1$, its asymptotic value is determined
by the global solution rather than fixed independently at infinity. As
discussed in Sec.~V, $c_z$ may subsequently be absorbed into a constant
rescaling of the longitudinal coordinate and therefore does not represent
an additional physical parameter.

For the full set of solutions considered, $N/\rho^2$, $K/\rho$, and
$\sigma$ approach their respective limiting values smoothly. In
particular, for the representative solution with $R_1=0.1$ the
longitudinal ratio approaches
\begin{equation}
c_z\simeq0.96869,
\label{eq:czrepresentative}
\end{equation}
rather than unity in the horizon-normalized coordinate used for the
integration. This is fully consistent with the asymptotic expansion
\[
K(\rho)=c_z\rho+\mathcal O(\rho^{-1}),
\]
which implies
\[
\frac{K(\rho)}{\rho}
=
c_z+\mathcal O(\rho^{-2}).
\]
If the standard Lemos longitudinal normalization is imposed instead,
the rescaled quantity $K/(c_z\rho)$ tends to unity for $\alpha=1$.

As a representative example, we consider the solution with
\[
R_1=0.1.
\]
The integration gives
\begin{equation}
\sigma_\infty=1.04907,
\qquad
c_z\simeq0.96869,
\qquad
R_\infty\simeq0.33,
\qquad
P_\infty\simeq-2.34,
\qquad
n_1\simeq-1.451.
\label{eq:representativedata}
\end{equation}

The asymptotic vortex amplitude is very close to its horizon value,
\[
P_0=-2.3469,
\]
showing that the function $P(\rho)$ evolves only weakly throughout the
exterior region. This near constancy does not imply that the
non-Abelian sector is dynamically trivial: the commutator contribution
$PR$ remains nonzero once the electric profile $R(\rho)$ develops away
from the horizon, and the asymptotic coefficients derived previously
retain explicit dependence on $P_\infty$ and its subleading falloff.

\begin{figure}[H]
\centering
\includegraphics[width=1.\columnwidth]{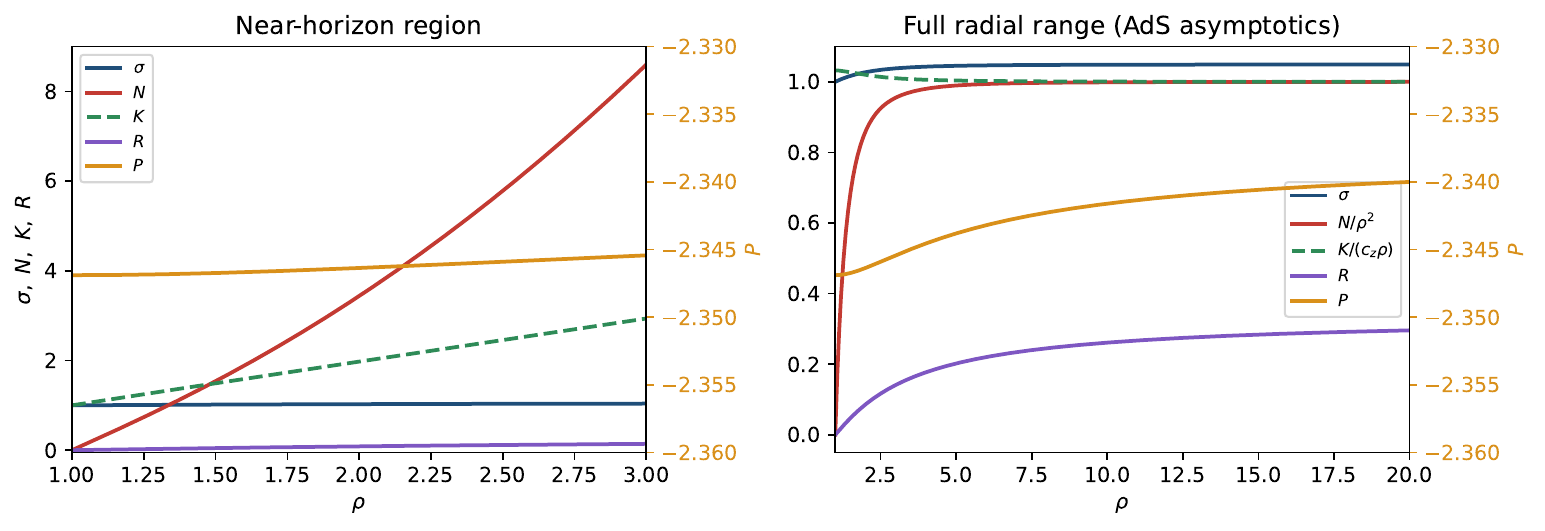}
\caption{Representative numerical solution for $R_1=0.1$. Left:
near-horizon region, $\rho\in[\rho_h,\rho_h+2]$. Right: extended radial
behavior, with the metric functions divided by their leading asymptotic
falloffs, $N/\rho^2\to1$ and $K/(c_z\rho)\to1$, so that they may be
displayed on the same scale as $\sigma$. For this solution,
$c_z\simeq0.96869$. In both panels, $P(\rho)$ is plotted against the
right-hand axis because its total variation is smaller than $1\%$ of its
horizon magnitude.}
\label{fig:profiles}
\end{figure}

Figure~\ref{fig:profiles} displays the representative solution in two
complementary radial ranges. The left panel focuses on the neighborhood
of the horizon. It confirms the leading behavior obtained analytically:
$N$ and $R$ grow linearly with $\rho-\rho_h$, while
$K-K_0$ also begins linearly. By contrast, the first derivative of the
vortex profile vanishes at the horizon, and its variation starts only at
quadratic order. Its total excursion over the interval shown is of order
\[
\Delta P\sim10^{-2},
\]
which explains why it would be nearly indistinguishable from a constant
if plotted on the same vertical scale as the remaining functions.

The right panel displays the approach to the asymptotic AdS region. The
rescaled functions $N/\rho^2$ and $K/(c_z\rho)$ both tend to unity, while
the lapse approaches the finite constant $\sigma_\infty$. The electric
potential $R(\rho)$ rises from zero at the horizon and approaches the
finite value $R_\infty$, whereas the vortex profile remains close to its
horizon value throughout the exterior. For clarity, the figure is shown
only up to $\rho=20$; the numerical integrations used to determine the
asymptotic quantities and perform the fits were extended to
$\rho=500$.

The representative solution therefore provides a smooth interpolation
between the regular near-horizon expansion and the asymptotic AdS
regime. More generally, the same qualitative behavior is observed
throughout the numerical branch explored in this work: the metric
functions remain regular outside the horizon, the electric potential
interpolates between $R(\rho_h)=0$ and a finite asymptotic value, and the
vortex profile evolves only mildly while continuing to influence both the
asymptotic lapse and the subleading metric coefficients. This overall
behavior is qualitatively consistent with the smooth connection between
regular horizons and AdS infinity found in other Einstein--Yang--Mills
black-hole families \cite{Bjoraker2000,Winstanley1999}, while exhibiting
the distinctive asymptotic features associated with the present
cylindrically symmetric non-Abelian vortex configuration. The dependence
of this solution branch on the color-electric parameter, including the
critical value at which the non-degenerate horizon expansion ceases to
exist, is examined below.

\subsection{Local horizon non-degeneracy bound}
\label{sec:extremality}

The family of solutions constructed in this work exists only while the
event horizon remains non-degenerate, which requires
\begin{equation}
N_1>0.
\end{equation}
As shown analytically in Sec.~IV, the leading horizon coefficient is
\begin{equation}
N_1
=
3\alpha^2\rho_h
-
\frac{4\pi G\rho_h}{s_0^2}R_1^2.
\label{eq:N1CriticalSection}
\end{equation}
The endpoint of the non-degenerate horizon expansion is therefore
determined exactly by the condition $N_1=0$, giving
\begin{equation}
R_1^{\rm crit}
=
s_0\alpha
\sqrt{\frac{3}{4\pi G}}.
\label{eq:Rcrit}
\end{equation}
In the numerical units
$G=\alpha=s_0=1$ used throughout Sec.~\ref{sec:numerics}, this becomes
\begin{equation}
R_1^{\rm crit}
=
\sqrt{\frac{3}{4\pi}}
=
0.488603\ldots .
\label{eq:RcritNumerical}
\end{equation}

The numerical horizon solutions independently reproduce this value. In
particular, the numerical root of $N_1(R_1)$ is
$R_1^{\rm crit}\simeq0.4886$, in agreement with
Eq.~\eqref{eq:RcritNumerical}. The absence of $\rho_h$ from
Eq.~\eqref{eq:Rcrit} also explains analytically the horizon-radius
independence of the critical value observed in the numerical solutions.

For
\begin{equation}
R_1>R_1^{\rm crit},
\end{equation}
Eq.~\eqref{eq:N1CriticalSection} gives $N_1<0$, and the regular
non-degenerate outer-horizon branch considered here therefore ceases to
exist. Thus $R_1^{\rm crit}$ should be interpreted as an exact local
non-degeneracy bound for the class of horizon expansions constructed in
this work.

Since $N_1$ controls both the leading zero of $N(\rho)$ and the surface
gravity, the limit $R_1\rightarrow R_1^{\rm crit}$ naturally suggests
an extremal configuration. Indeed, the Hawking temperature derived in
Sec.~\ref{sec:thermo} tends to zero as $N_1\rightarrow0$. Nevertheless,
we deliberately refrain from identifying
Eq.~\eqref{eq:Rcrit} with a genuine global extremality condition.
The horizon expansion used throughout this work assumes a simple zero,
\begin{equation}
N(\rho)
=
N_1(\rho-\rho_h)+\cdots,
\end{equation}
and therefore ceases to be the appropriate local expansion precisely at
$N_1=0$. Establishing the existence of an extremal black string requires
a separate degenerate-horizon analysis beginning instead with
\begin{equation}
N(\rho)
=
N_2(\rho-\rho_h)^2+\cdots,
\end{equation}
together with a demonstration that the resulting local solution extends
regularly to the AdS boundary. This problem lies beyond the scope of the
present work.

\section{Abelian Limit and the Onset of Non-Abelian Hair}
\label{sec:analyticweak}

The numerical solutions constructed in Sec.~\ref{sec:numerics} belong to
the genuinely non-Abelian sector of the $SU(2)$ gauge ansatz
\begin{equation}
A
=
\tau_2R(\rho)\,dt
+
\tau_3P(\rho)\,d\varphi,
\label{eq:gaugeAnsatzLimits}
\end{equation}
since both gauge functions are nonvanishing. In particular, the field
strength contains the commutator contribution
\begin{equation}
F^{(1)}_{t\varphi}
=
PR,
\label{eq:commutatorDiagnostic}
\end{equation}
which is absent in a single-generator Abelian embedding.

The ansatz contains two physically distinct limiting sectors. They are
useful both for interpreting the numerical solutions and for identifying
the origin of the non-Abelian branch.

\subsection{Neutral and Abelian limits}

First consider the neutral endpoint. Regularity at a non-degenerate
horizon requires
\begin{equation}
R(\rho_h)=0.
\end{equation}
If in addition
\begin{equation}
R_1\equiv R'(\rho_h)=0,
\label{eq:R1NeutralLimit}
\end{equation}
the electric Yang--Mills equation,
\begin{equation}
\left(
\frac{K\rho}{\sigma}R'
\right)'
=
\frac{KP^2R}{\rho N\sigma},
\end{equation}
implies, for a regular analytic horizon expansion, that
\begin{equation}
R(\rho)\equiv0.
\label{eq:RNeutralExact}
\end{equation}

The remaining Yang--Mills equation then becomes
\begin{equation}
\left(
\frac{\sigma KN}{\rho}P'
\right)'=0.
\end{equation}
After one integration,
\begin{equation}
\frac{\sigma KN}{\rho}P'=C_P.
\end{equation}
Since $N\sim N_1(\rho-\rho_h)$ at the horizon, a nonzero $C_P$ would
produce $P'\sim(\rho-\rho_h)^{-1}$ and hence a logarithmically divergent
gauge profile. Regularity therefore requires $C_P=0$, so that
\begin{equation}
P(\rho)=P_\star={\rm const.}
\end{equation}
throughout the exterior region.

Consequently,
\begin{equation}
F^a_{\mu\nu}=0,
\qquad
T^{\rm YM}_{\mu\nu}=0.
\end{equation}
The geometry therefore reduces exactly to the neutral Lemos black string.
In the canonical longitudinal normalization it may be written as
\begin{equation}
\sigma=1,
\qquad
K=\alpha\rho,
\end{equation}
with
\begin{equation}
N_L(\rho)
=
\alpha^2\rho^2
-
\frac{\alpha^2\rho_h^3}{\rho}.
\label{eq:neutralLemosLimit}
\end{equation}
Thus the configuration at $R_1=0$ is exactly the neutral Lemos solution
with vanishing Yang--Mills field strength. The choice $K=\alpha\rho$
corresponds here to the conventional normalization of the longitudinal
coordinate. As discussed in Secs.~V and~\ref{sec:numerics}, the
horizon-normalized coordinate employed in the numerical construction
generically yields $K(\rho)\sim c_z\rho$ at infinity; the constant $c_z$
can be absorbed by a constant rescaling of the longitudinal coordinate
and does not represent an additional physical degree of freedom.

There is, however, a second and physically different limit. Setting
\begin{equation}
P(\rho)=0,
\qquad
R(\rho)\neq0,
\label{eq:AbelianLimitFinal}
\end{equation}
restricts the gauge potential to the single internal direction $\tau_2$.
All commutator terms then vanish, and the $SU(2)$ configuration reduces
to an Abelian $U(1)\subset SU(2)$ embedding.

Choosing again the canonical longitudinal normalization, the Abelian
background may be written as
\begin{equation}
\sigma_0=1,
\qquad
K_0=\alpha\rho.
\end{equation}
This is related to the horizon-normalized coordinate used for the
non-Abelian numerical solutions by a constant rescaling of the
longitudinal coordinate. The electric Yang--Mills equation then gives
\begin{equation}
R_0(\rho)
=
\mu-\frac{Q}{\rho}.
\end{equation}
Choosing the gauge in which the electric potential vanishes at the
horizon,
\begin{equation}
R_0(\rho_h)=0,
\end{equation}
fixes
\begin{equation}
\mu=\frac{Q}{\rho_h},
\end{equation}
and therefore
\begin{equation}
R_0(\rho)
=
Q\left(
\frac{1}{\rho_h}
-
\frac{1}{\rho}
\right).
\label{eq:RAbelianBackground}
\end{equation}

For the normalization of the Yang--Mills action adopted in this work,
the corresponding Einstein equations yield
\begin{equation}
N_0(\rho)
=
\alpha^2\rho^2
-
\frac{M}{\rho}
+
\frac{4\pi GQ^2}{\rho^2}.
\label{eq:NAbelianBackground}
\end{equation}
The horizon condition $N_0(\rho_h)=0$ fixes
\begin{equation}
M
=
\alpha^2\rho_h^3
+
\frac{4\pi GQ^2}{\rho_h}.
\label{eq:MAbelianBackground}
\end{equation}

The normalization in Eq.~\eqref{eq:NAbelianBackground} follows directly
from the convention
\begin{equation}
R_0'(\rho)=\frac{Q}{\rho^2}.
\end{equation}
At the horizon,
\begin{equation}
R_1
=
R_0'(\rho_h)
=
\frac{Q}{\rho_h^2},
\label{eq:QandR1}
\end{equation}
and therefore
\begin{equation}
N_0'(\rho_h)
=
3\alpha^2\rho_h
-
4\pi G\rho_h R_1^2.
\label{eq:N1AbelianR1}
\end{equation}
For the horizon normalization $s_0=1$, this is precisely the expression
obtained independently for the full non-Abelian branch in
Eq.~\eqref{eq:N1horizonExact}. Thus the leading coefficient $N_1$ of the
horizon function is insensitive to the transverse non-Abelian hair,
despite the fact that the complete solutions are not Abelian. As shown
below, this agreement reflects the perturbative structure of the
Yang--Mills stress tensor: the transverse non-Abelian mode appears at
linear order in the gauge field, whereas its contribution to the
gravitational backreaction begins at quadratic order.

The two limits should therefore be distinguished clearly:
\begin{align}
R=0,\qquad P=P_\star
&\quad\Longrightarrow\quad
F^a_{\mu\nu}=0
\quad\text{(neutral Lemos)},                       \label{eq:twoLimitsA}\\
P=0,\qquad R\neq0
&\quad\Longrightarrow\quad
U(1)\subset SU(2)
\quad\text{(charged Abelian black string)}.        \label{eq:twoLimitsB}
\end{align}

\subsection{Linear onset of the non-Abelian branch}
\label{sec:nonAbelianOnset}

The charged Abelian solution provides a natural analytic background from
which to study the onset of the genuinely non-Abelian branch. We adopt
the same canonical longitudinal normalization for the perturbative
analysis and introduce a small parameter $\varepsilon$ measuring the
amplitude of the gauge component transverse to the Abelian embedding,
\begin{equation}
P(\rho)
=
\varepsilon\,\psi(\rho)
+
\mathcal O(\varepsilon^3).
\label{eq:POnset}
\end{equation}

Keeping the Abelian mass and electric charge fixed, the remaining fields
may be expanded as
\begin{align}
R(\rho)
&=
R_0(\rho)
+
\varepsilon^2r_2(\rho)
+
\mathcal O(\varepsilon^4),
\label{eq:ROnset}
\\
N(\rho)
&=
N_0(\rho)
+
\varepsilon^2n_2(\rho)
+
\mathcal O(\varepsilon^4),
\nonumber\\
\sigma(\rho)
&=
1+
\varepsilon^2s_2(\rho)
+
\mathcal O(\varepsilon^4),
\nonumber\\
K(\rho)
&=
\alpha\rho
\left[
1+\varepsilon^2k_2(\rho)
+\mathcal O(\varepsilon^4)
\right].
\label{eq:metricOnset}
\end{align}
In this normalization, any constant asymptotic contribution carried by
$k_2$ can be removed order by order by a constant rescaling of the
longitudinal coordinate. This is the perturbative counterpart of the
finite rescaling that converts the numerical asymptotic behavior
$K(\rho)\sim c_z\rho$ into the canonical AdS normalization.

The absence of linear corrections in
Eqs.~\eqref{eq:ROnset}--\eqref{eq:metricOnset} follows directly from the
internal structure of the Yang--Mills field. The Abelian background has
only
\begin{equation}
F^{(0)(2)}_{t\rho}
=
-R_0',
\end{equation}
whereas the perturbation~\eqref{eq:POnset} generates
\begin{equation}
\delta F^{(3)}_{\rho\varphi}
=
\varepsilon\psi',
\qquad
\delta F^{(1)}_{t\varphi}
=
\varepsilon R_0\psi.
\label{eq:linearNAFieldStrength}
\end{equation}
The latter component is entirely due to the commutator
$[A_t,A_\varphi]$ and therefore provides a direct signature of the
non-Abelian character of the perturbation.

The Yang--Mills stress tensor is built from contractions over equal
internal indices,
\begin{equation}
T^{\rm YM}_{\mu\nu}
\propto
\sum_a
F^a_{\mu\lambda}F^{a\lambda}{}_{\nu}
+\cdots .
\end{equation}
At $\mathcal O(\varepsilon)$, the field strength splits as
$F=F^{(0)}+\delta F$, where the Abelian background carries only the
internal index $a=2$, while the transverse perturbation in
Eq.~\eqref{eq:linearNAFieldStrength} carries only $a=1,3$. Since the
Killing form is diagonal in this basis,
$\langle\tau_a,\tau_b\rangle\propto\delta_{ab}$, the cross term between
the Abelian background and the transverse linear perturbation vanishes.
Consequently,
\begin{equation}
T^{\rm YM}_{\mu\nu}
=
T^{(0)}_{\mu\nu}
+
\mathcal O(\varepsilon^2),
\label{eq:TNoLinear}
\end{equation}
and there is no genuinely non-Abelian metric correction at
$\mathcal O(\varepsilon)$.

The electric Yang--Mills equation leads to the same order counting. Its
non-Abelian source is proportional to $P^2R$, so that the first
deformation of the Abelian electric profile occurs at
$\mathcal O(\varepsilon^2)$.

At linear order, therefore, the complete non-Abelian problem reduces to
a single equation for the transverse mode,
\begin{equation}
\left(
N_0\psi'
\right)'
+
\frac{R_0^2}{N_0}\psi
=
0.
\label{eq:NAOnsetEquation}
\end{equation}
This equation describes the infinitesimal departure from the charged
$U(1)$-embedded black string toward the genuinely non-Abelian $SU(2)$
sector.

Regularity at a non-degenerate horizon follows directly from
Eq.~\eqref{eq:NAOnsetEquation}. Since
\begin{equation}
N_0
=
N_{01}(\rho-\rho_h)+\cdots,
\qquad
R_0
=
R_1(\rho-\rho_h)+\cdots,
\end{equation}
substitution into Eq.~\eqref{eq:NAOnsetEquation} gives at leading order
\begin{equation}
N_{01}\psi'(\rho_h)=0.
\end{equation}
For a non-degenerate horizon, $N_{01}\neq0$, and therefore
\begin{equation}
\psi'(\rho_h)=0.
\label{eq:PsiHorizon}
\end{equation}
The value $\psi(\rho_h)$ consequently sets the amplitude of the
infinitesimal non-Abelian deformation.

For any nontrivial solution with $\psi\neq0$,
\begin{equation}
F^{(1)}_{t\varphi}
=
\varepsilon R_0\psi
+
\mathcal O(\varepsilon^3)
\end{equation}
is nonzero away from the horizon. Hence the perturbation immediately
leaves the Abelian single-generator sector, even though the geometry
remains equal to the charged Abelian background at linear order. In this
sense, Eq.~\eqref{eq:NAOnsetEquation} determines a transverse direction
from the embedded $U(1)$ subfamily into the genuinely non-Abelian sector
of the $SU(2)$ theory.

\subsection{Quadratic backreaction}

The perturbative hierarchy is therefore
\begin{equation}
P=\mathcal O(\varepsilon),
\qquad
R-R_0=\mathcal O(\varepsilon^2),
\qquad
g_{\mu\nu}-g^{(0)}_{\mu\nu}
=
\mathcal O(\varepsilon^2).
\label{eq:NAHierarchy}
\end{equation}
The Yang--Mills configuration thus becomes genuinely non-Abelian before
its gravitational backreaction appears.

The first correction to the electric gauge function illustrates
explicitly how the transverse mode feeds back into the original Abelian
sector. Expanding the electric Yang--Mills equation to quadratic order
gives
\begin{equation}
\left\{
\rho^2
\left[
r_2'
+
(k_2-s_2)R_0'
\right]
\right\}'
=
\frac{R_0}{N_0}\psi^2.
\label{eq:RQuadraticSource}
\end{equation}
Thus the transverse mode determined by
Eq.~\eqref{eq:NAOnsetEquation} acts as a source for the deformation of
the electric profile at the next order.

The same quadratic combinations of $\psi$ and $\psi'$ source the
Einstein equations for $n_2$, $s_2$, and $k_2$. Once the linear mode
$\psi$ is known, the second-order equations form a linear inhomogeneous
system for these corrections. We do not pursue its explicit integration
here, since the main purpose of the present analysis is to identify the
analytic origin and perturbative order structure of the non-Abelian
branch rather than to replace the fully backreacted numerical solutions.

The resulting picture is therefore simple. The charged Abelian
Lemos--Zanchin solutions form an exact embedded $U(1)$ subfamily of the
$SU(2)$ theory. Equation~\eqref{eq:NAOnsetEquation} determines a
transverse Yang--Mills mode that takes the system away from this Abelian
subfamily. The gauge configuration becomes genuinely non-Abelian already
at $\mathcal O(\varepsilon)$ through the commutator component
$F^{(1)}_{t\varphi}$, whereas the original electric profile and the
geometry respond only at $\mathcal O(\varepsilon^2)$. Schematically,
\begin{equation}
\text{charged Abelian black string}
\;\xrightarrow{\;\mathcal O(\varepsilon)\;}
\text{non-Abelian Yang--Mills hair}
\;\xrightarrow{\;\mathcal O(\varepsilon^2)\;}
\text{gravitational backreaction}.
\label{eq:AbelianToNonAbelian}
\end{equation}

The fully backreacted numerical solutions of Sec.~\ref{sec:numerics} may
therefore be naturally interpreted as nonlinear realizations of this
transverse non-Abelian deformation. Strictly speaking, however,
establishing a genuine global bifurcation from the embedded Abelian family
would require solving the linearized boundary-value
problem~\eqref{eq:NAOnsetEquation} throughout the complete exterior region
and verifying the corresponding AdS asymptotic condition for
$\psi(\rho)$. Such a global linearized analysis lies beyond the scope of
the present work. Together with the neutral endpoint discussed above,
this identifies two distinguished limiting sectors of the solution
family: the neutral Lemos geometry obtained when the Yang--Mills field
strength vanishes, and the charged Abelian black string obtained when the
transverse non-Abelian component is switched off.
\section{Thermodynamics}
\label{sec:thermo}

The thermodynamic properties of black holes and black strings are
governed by the laws of black-hole mechanics and the Hawking effect
\cite{Bardeen1973,Hawking1975}, and, in the present case, follow directly
from the horizon expansion together with the asymptotic normalization of
the time coordinate. Unlike the standard one-function
Lemos geometry, the present three-function metric possesses an
independent asymptotic lapse $\sigma_\infty$, which must be taken into
account when defining the physical surface gravity.

For the metric~(\ref{eq:metric}), the surface gravity is
\[
\kappa
=
\frac12\sqrt{A'(\rho_h)B'(\rho_h)},
\]
with
\[
A=\sigma^2N,
\qquad
B=N.
\]

Using $N(\rho_h)=0$, one obtains
\[
\kappa
=
\frac12 s_0N_1.
\]
This expression, however, corresponds to the horizon normalization
$s_0=1$ adopted for the numerical integration. The physical Hawking
temperature must instead be referred to the asymptotic observer, for whom
the time coordinate is normalized by
\[
t\rightarrow\sigma_\infty t,
\]
so that the metric approaches pure AdS at infinity. Consequently,
\begin{equation}
T_H
=
\frac{s_0N_1}
{4\pi\sigma_\infty}.
\label{eq:TH}
\end{equation}

The entropy follows the Bekenstein--Hawking area law
\cite{Bekenstein1973,Hawking1975},
\begin{equation}
S
=
\frac{A_h}{4G}
=
\frac{\pi K_0\rho_h\Delta z}
{2G},
\label{eq:entropy}
\end{equation}
where $\Delta z$ denotes the (possibly compactified) coordinate length of
the string in the horizon-normalized coordinate used in the numerical
integration. Throughout the charge scan discussed below,
$\rho_h=K_0=1$, and therefore the entropy per unit coordinate length in
this normalization is constant,
\begin{equation}
\frac{S}{\Delta z}
=
\frac{\pi}{2}.
\label{eq:entropyCoordinate}
\end{equation}
This coordinate entropy density should be distinguished from the entropy
per unit canonically normalized asymptotic length. As discussed in
Secs.~V--VII, the numerical solutions behave as
$K(\rho)\sim c_z\rho$ at large radius. Introducing the canonical
longitudinal coordinate
\begin{equation}
\bar z=\frac{c_z}{\alpha}z,
\end{equation}
one obtains
\begin{equation}
\frac{S}{\Delta\bar z}
=
\frac{\pi\alpha K_0\rho_h}{2Gc_z}.
\label{eq:entropyCanonical}
\end{equation}
Thus the total Bekenstein--Hawking entropy is invariant under the constant
longitudinal rescaling, as expected, whereas its density per unit
coordinate length depends on the normalization adopted.

The resulting numerical values of the horizon coefficient $N_1$, the
asymptotic lapse $\sigma_\infty$, and the corresponding Hawking
temperature are listed in Table~\ref{tab:thermo}.
\begin{table}[h]
\centering
\begin{tabular}{cccc}
\hline\hline
$R_1$ & $N_1$ & $\sigma_\infty$ & $T_H$ \\
\hline
0.05 & 2.9686 & 1.0126 & 0.2333 \\
0.10 & 2.8743 & 1.0491 & 0.2180 \\
0.15 & 2.7173 & 1.1047 & 0.1957 \\
0.20 & 2.4973 & 1.1707 & 0.1698 \\
0.25 & 2.2146 & 1.2330 & 0.1429 \\
0.30 & 1.8690 & 1.2743 & 0.1167 \\
0.35 & 1.4606 & 1.2772 & 0.0910 \\
0.40 & 0.9894 & 1.2297 & 0.0640 \\
0.45 & 0.4553 & 1.1258 & 0.0322 \\
\hline\hline
\end{tabular}
\caption{$N_1$ (the closed-form, horizon-only coefficient of Sec.~IV),
the asymptotic lapse $\sigma_\infty$ (requiring full integration to
infinity), and the resulting physical Hawking temperature $T_H$
(Eq.~\eqref{eq:TH}), as functions of charge. In the horizon-normalized
integration coordinate, $S/\Delta z=\pi/2$ throughout, since
$\rho_h=K_0=1$ are held fixed. $T_H$ approaches zero as the exact local
horizon non-degeneracy bound
$R_1^{\rm crit}=\sqrt{3/(4\pi)}\approx0.4886$ is approached.}
\label{tab:thermo}
\end{table}

The table immediately reveals two distinct features.
First, the asymptotic lapse is not constant: starting from values close to
unity at small charge, it increases to a maximum around
$R_1\simeq0.35$ before decreasing again as the critical horizon bound is
approached. Consequently, the Hawking temperature is not simply
proportional to the horizon coefficient $N_1$, despite the latter being
given analytically by the horizon expansion.
Instead, the complete radial integration is essential for obtaining the
physical temperature because the normalization of the time coordinate is
fixed only at infinity.

Second, the entropy per unit coordinate length in the horizon-normalized
integration gauge remains constant along the present charge scan because
the horizon data $\rho_h$ and $K_0$ are held fixed. When expressed per
unit canonically normalized asymptotic length, however, the corresponding
entropy density acquires the factor $1/c_z$, as shown in
Eq.~\eqref{eq:entropyCanonical}. The Hawking temperature, by contrast,
varies nontrivially through the asymptotic lapse $\sigma_\infty$.
Thus, within the family compared at fixed horizon data
$(\rho_h,K_0)$ in the integration gauge, the temperature provides a
thermodynamic diagnostic of the non-Abelian sector.
\begin{figure}[b]
\centering
\includegraphics[width=1.\columnwidth]{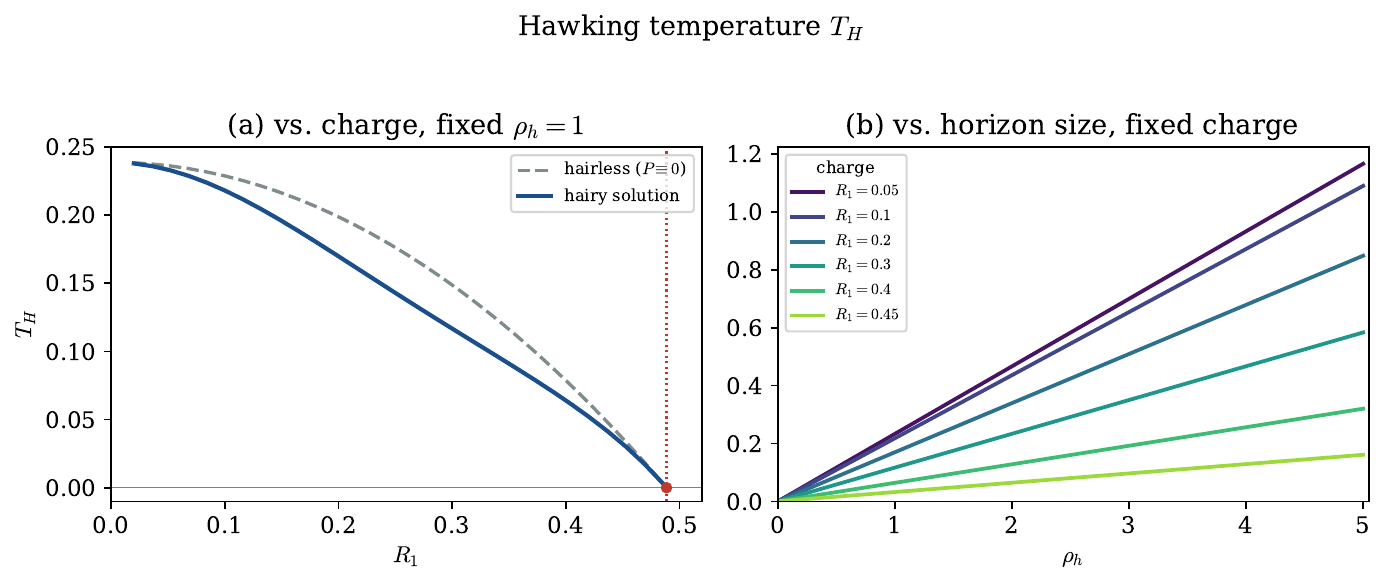}
\caption{Hawking temperature as a function of the color-electric parameter
(panel a) and the horizon radius (panel b), computed from
Eq.~\eqref{eq:TH}. In panel (a), the present Einstein--Yang--Mills black
string (solid line) is compared with the corresponding hairless
Lemos-–Zanchin solution (dashed line) for
$\rho_h=\alpha=s_0=K_0=G=1$. In panel (b), the temperature is shown as a
function of $\rho_h$ for representative subcritical values of
$R_1<R_1^{\rm crit}$, with $\alpha=s_0=K_0=G=1$.}
\label{fig:temperature}
\end{figure}
Figure~\ref{fig:temperature}(a) compares the present non-Abelian solution
with the corresponding embedded Abelian configuration obtained by setting
$P(\rho)\equiv0$. In this limit the gauge field is restricted to a single
internal generator, the metric reduces to the charged Lemos--Zanchin form
\cite{Lemos1995,LemosZanchin1996}, and
\begin{equation}
\sigma(\rho)\equiv1.
\end{equation}
As shown analytically in Eqs.~\eqref{eq:N1horizonExact} and
\eqref{eq:N1AbelianR1}, the leading horizon coefficient $N_1$ is exactly
the same for the non-Abelian and embedded Abelian solutions when they are
compared at fixed $(\rho_h,R_1)$. The difference in their Hawking
temperatures therefore arises from the asymptotic normalization of the
time coordinate. For the non-Abelian branch,
$\sigma_\infty$ is determined by the fully backreacted solution, whereas
the embedded Abelian solution has $\sigma_\infty=1$. Consequently,
\begin{equation}
\frac{T_H^{\rm NA}}{T_H^{\rm Abelian}}
=
\frac{1}{\sigma_\infty},
\label{eq:temperatureRatio}
\end{equation}
at fixed $(\rho_h,R_1)$. In the range for which
$\sigma_\infty>1$ along the numerical branch, this gives
\begin{equation}
T_H^{\rm NA}
<
T_H^{\rm Abelian}.
\end{equation}

This inequality refers specifically to the comparison at fixed
$(\rho_h,R_1)$. Since neither $\rho_h$ nor $R_1$ is, by itself, a
conserved asymptotic charge, it should not be interpreted as a comparison
between the two branches at fixed physical mass and charge. Such a
comparison would require an independent determination of the conserved
asymptotic charges. Within the fixed-$(\rho_h,R_1)$ comparison adopted
here, however, a clear hierarchy emerges among the physical observables.
At fixed $(\rho_h,R_1)$, the leading horizon coefficient $N_1$
coincides with that of the embedded Abelian solution. The horizon area
per unit integration-coordinate length is likewise fixed by construction,
although its density with respect to the canonically normalized
asymptotic longitudinal coordinate carries the additional factor
$1/c_z$. The distinction between the two branches therefore appears
thermodynamically through global asymptotic normalization factors:
$\sigma_\infty$ controls the Hawking temperature, while $c_z$ controls
the conversion between coordinate and canonically normalized longitudinal
entropy densities. In particular, the temperature shift produced by the
non-Abelian hair is a genuinely global effect: it is not encoded in the
leading local horizon coefficient, but emerges only after the solution is
continued from the horizon to the AdS boundary.

Figure~\ref{fig:temperature}(b) illustrates a second consequence of the
scaling relations established in Sec.~IV.
Since $\sigma_\infty$ was found numerically to be independent of
$\rho_h$ within the explored branch, the temperature inherits the same
linear dependence on the horizon radius as $N_1$,
\[
T_H
=
\frac{\hat N_1(R_1)}
{4\pi\sigma_\infty(R_1)}
\,\rho_h,
\]
for every fixed subcritical value of $R_1$.
None of these curves crosses the horizontal axis; instead, the vanishing
of the temperature occurs only when the charge approaches the critical
value identified in the previous section.

Finally, the thermodynamic signature identified here should be clearly
distinguished from the asymptotic geometric signature discussed in
Sec.~V. The reduction of the Hawking temperature relative to the
Lemos-–Zanchin solution originates entirely from the asymptotic lapse
normalization and is therefore encoded in the single quantity
$\sigma_\infty$. By contrast, the genuinely irreducible imprint of the
non-Abelian sector on the spacetime geometry remains the cubic
coefficient $n_3$, which cannot be absorbed into a redefinition of an
effective Abelian charge. The present solution thus exhibits two
complementary manifestations of the Yang--Mills hair: a thermodynamic one,
through the global normalization of the temperature, and a geometric one,
through the asymptotic structure of the metric.
\section{Final Remarks}
\label{sec:conclusions}

In this work we have constructed a static, cylindrically symmetric,
electrically charged $SU(2)$ Einstein--Yang--Mills black string in
asymptotically anti-de Sitter spacetime. To our knowledge, this provides
the first black-string realization of the electric counterpart of the
Gal'tsov--Davydov--Volkov vortex configuration and, for the linear
Yang--Mills theory, completes the two extensions explicitly identified in
their original work: the introduction of an event horizon and the
$t\leftrightarrow z$ electric generalization of the magnetic vortex
ansatz. We have constructed only the exterior region $\rho\ge\rho_h$; whether the interior admits an inner (Cauchy) horizon, or whether the solution is regular at the origin, remains entirely open, to be explored in a future work.

A central result of the present analysis is that the standard
single-function cylindrical black-string geometry is incompatible with a
genuinely non-Abelian vortex source whenever the electric and vortex
sectors coexist. A consistent description requires a metric with three
independent radial functions, from which we derived the complete reduced
Einstein--Yang--Mills system and its regular near-horizon and asymptotic
AdS expansions. The numerical integration of this system yields a smooth
family of black-string solutions interpolating between a regular event
horizon and the asymptotic anti-de Sitter region.

The resulting solutions exhibit several characteristic features. At the
horizon, the nonlinear algebraic system numerically selects a definite
value of the vortex profile for each pair $(\rho_h,R_1)$, rather than
leaving it as an independent continuous shooting parameter, although a
complete Jacobian analysis will be required to establish this as a
rigorous local uniqueness result. At the same time, the leading horizon
coefficient is obtained analytically as
\begin{equation}
N_1
=
3\alpha^2\rho_h
-
\frac{4\pi G\rho_h}{s_0^2}R_1^2,
\end{equation}
which yields the exact local non-degeneracy bound
\begin{equation}
R_1^{\rm crit}
=
s_0\alpha\sqrt{\frac{3}{4\pi G}}.
\end{equation}
This result explains analytically both the horizon-radius independence of
the critical value and the numerical result
$R_1^{\rm crit}\simeq0.4886$ in the units adopted in this work.

The gauge ansatz also contains two distinguished limiting sectors. When
the electric horizon datum is switched off, regularity forces the
Yang--Mills field strength to vanish and the neutral Lemos black string is
recovered exactly. By contrast, setting the transverse vortex component
to zero yields an embedded $U(1)\subset SU(2)$ sector described by the
charged Lemos--Zanchin black string. Perturbing about this exact Abelian
background shows that a transverse gauge mode generates genuinely
non-Abelian hair already at linear order through the commutator component
$F^{(1)}_{t\varphi}$, while the electric profile and the spacetime geometry
respond only at quadratic order in the transverse amplitude. The numerical
solutions constructed here may therefore be naturally interpreted as fully
backreacted nonlinear realizations associated with this transverse
deformation away from the embedded Abelian subfamily.

In the asymptotic region, the metric develops two qualitatively distinct
subleading corrections: the coefficient $n_2$, which can still be
reinterpreted in terms of an effective Abelian charge, and the cubic
coefficient $n_3$, whose dependence on the asymptotic vortex amplitude
and its subleading falloff constitutes the first irreducible geometric
signature of the non-Abelian hair within the comparison class considered
here. Thermodynamically, the entropy obeys the usual area law, with its density per unit canonically normalized longitudinal length differing from that in the horizon-normalized integration coordinate by the corresponding factor involving $c_z$, whereas the Hawking temperature distinguishes the non-Abelian and embedded Abelian branches through the asymptotic lapse normalization.
At fixed $(\rho_h,R_1)$ their leading horizon coefficient $N_1$ is
identical, but the non-Abelian solution develops a nontrivial
$\sigma_\infty$, so that
\begin{equation}
\frac{T_H^{\rm NA}}{T_H^{\rm Abelian}}
=
\frac{1}{\sigma_\infty}.
\end{equation}
The thermal effect of the non-Abelian hair is therefore global rather
than encoded in the leading local horizon coefficient.

The present results expand the class of known cylindrically symmetric Einstein--Yang--Mills configurations and show that non-Abelian vortex hair produces distinctive physical effects extending from the local horizon geometry to the asymptotic structure and thermal properties of the spacetime. More generally, they demonstrate that the cylindrical sector possesses a substantially richer solution space than
suggested by the familiar Abelian black-string geometries.

Several natural directions emerge from the present work. A first priority
is the explicit construction of the degenerate branch associated with the
exact local bound $N_1=0$. Since the non-degenerate expansion employed
here assumes a simple zero of $N(\rho)$, the critical point requires a
separate near-horizon analysis beginning with
$N(\rho)=N_2(\rho-\rho_h)^2+\cdots$. Such a construction would determine
whether the exact local bound
$R_1^{\rm crit}=s_0\alpha\sqrt{3/(4\pi G)}$ corresponds to a genuine
extremal black string in the global sense. Another important step is the
determination of the conserved physical mass through an asymptotic prescription, such as the Brown--York or holographic stress-tensor formalism, providing the normalization needed
for a complete thermodynamic description. This would allow a systematic
investigation of the first law, a possible Smarr relation, and derived
thermodynamic quantities such as the heat capacity and Helmholtz free
energy. A complementary direction is the analysis of dynamical stability.
Although asymptotically flat colored black holes are generically affected
by unstable modes, asymptotically AdS Einstein--Yang--Mills systems are
known to admit dynamically stable sectors for suitable gauge-field
configurations \cite{Winstanley1999,Bjoraker2000,Baxter2008}. It would
therefore be important to determine whether the cylindrical non-Abelian
branch constructed here possesses an analogous stability window. Finally, extending the present
construction to nonlinear Yang--Mills theories, especially
Born--Infeld-type models, offers a natural opportunity to explore how
non-Abelian nonlinearities modify the horizon structure, asymptotic
geometry, and thermodynamic behavior of cylindrically symmetric black
strings.

\begin{acknowledgments}
\noindent The authors would like to thank the Conselho Nacional de Desenvolvimento Cient\'{i}fico e Tecnol\'{o}gico (CNPq) for partial financial support. 
\end{acknowledgments}

\bibliographystyle{apsrev4-1}
\bibliography{Ref1}

\appendix

\section{Reduced Einstein--Yang--Mills Equations}
\label{app:equations}

For completeness and reproducibility, we collect here the explicit
Einstein--Yang--Mills equations underlying the reduced system discussed in
Sec.~III.

For the linear Yang--Mills Lagrangian,
$\mathcal L=-\mathcal F/4$, and the metric
ansatz~\eqref{eq:metric}, the reduced matter Lagrangian becomes
\begin{equation}
\mathcal L_{\rm red}
=
\frac{K\rho}{2\sigma}R'^2
-
\frac{\sigma KN}{2\rho}P'^2
+
\frac{KP^2R^2}{2\rho N\sigma},
\end{equation}
where $\sqrt{-g}=\sigma\rho K$.

Variation with respect to the Yang--Mills functions gives
\begin{align}
\left(
\frac{K\rho}{\sigma}R'
\right)'
&=
\frac{KP^2R}{\rho N\sigma},
\label{eq:appELR}
\\
\left(
\frac{\sigma KN}{\rho}P'
\right)'
&=
-
\frac{KPR^2}{\rho N\sigma}.
\label{eq:appELP}
\end{align}

The corresponding non-vanishing diagonal components of the Einstein
tensor are
\begin{align}
G^t_{\ t}
&=
\frac{
KN'
+
2K'N
+
K'N'\rho
+
2K''N\rho
}{
2K\rho
},
\\
G^\rho_{\ \rho}
&=
\frac{
2KN\sigma'
+
KN'\sigma
+
2K'N\rho\sigma'
+
2K'N\sigma
+
K'N'\rho\sigma
}{
2K\rho\sigma
},
\\
G^\varphi_{\ \varphi}
&=
\frac{
2KN\sigma''
+
3KN'\sigma'
+
KN''\sigma
+
2K'N\sigma'
+
2K'N'\sigma
+
2K''N\sigma
}{
2K\sigma
},
\\
G^z_{\ z}
&=
\frac{
2N\rho\sigma''
+
2N\sigma'
+
3N'\rho\sigma'
+
2N'\sigma
+
N''\rho\sigma
}{
2\rho\sigma
}.
\end{align}
These quantities satisfy
\[
G^\mu_{\ \nu}
+
\Lambda\delta^\mu_{\ \nu}
=
8\pi G
T^\mu_{\ \nu},
\]
together with
Eqs.~\eqref{eq:appELR}--\eqref{eq:appELP}.

As discussed in Sec.~III, only three of the four diagonal Einstein
equations are independent owing to the contracted Bianchi identity.
The numerical integrations reported throughout this work were performed
using the reduced first-order system obtained from these equations, while
every solution was subsequently verified against the complete Einstein and
Yang--Mills equations, yielding residuals at the level of machine
precision.

\section{Structure of the Horizon System}
\label{app:horizon}

The horizon expansion discussed in Sec.~IV possesses a subtle structural
feature that deserves additional comment. Although the Einstein and
Yang--Mills equations are fully coupled, the non-Abelian interaction enters
the two sectors at different orders of the near-horizon expansion. This
difference allows the color-electric datum and the vortex amplitude to play
structurally distinct roles in the local boundary-value problem.

Near the event horizon, the regular Taylor expansions begin as
\begin{equation}
R(\rho)=R_1(\rho-\rho_h)+\cdots,
\qquad
N(\rho)=N_1(\rho-\rho_h)+\cdots,
\label{eq:horA1}
\end{equation}
with analogous expansions for $\sigma(\rho)$, $K(\rho)$, and
$P(\rho)$.

Substituting Eq.~\eqref{eq:horA1} into the Yang--Mills equation
\eqref{eq:appELR} gives
\begin{equation}
\frac{KP^2R}{\rho N\sigma}
\longrightarrow
\frac{K_0P_0^{\,2}R_1}
     {N_1\rho_h s_0},
\label{eq:horA2}
\end{equation}
which remains finite at the horizon. Consequently, the vortex amplitude
$P_0$ already contributes to the leading-order algebraic horizon system.

The corresponding non-Abelian contribution to the Einstein equations
behaves instead as
\begin{equation}
\frac{P^2R^2}
     {\rho^2N\sigma^2}
=
\mathcal O(\rho-\rho_h),
\label{eq:horA3}
\end{equation}
because it is quadratic in the electric profile $R(\rho)$. It therefore
enters only at the next order of the expansion, contributing to the
equations that determine the coefficients $N_2$, $K_1$, and $s_1$.

The different order counting expressed by
Eqs.~\eqref{eq:horA2} and \eqref{eq:horA3} is the mechanism underlying
the structure of the horizon problem. The leading metric coefficient
$N_1$ is determined through the Einstein sector by the color-electric
datum $R_1$, whereas the vortex amplitude $P_0$ already enters the
leading-order Yang--Mills condition. Thus, $R_1$ and $P_0$ enter the
near-horizon problem in structurally distinct ways, even though the
electric and vortex sectors remain fully coupled in the complete field
equations.

A natural question is whether the resulting horizon system determines a
locally unique value of $P_0$. To investigate whether the numerical
determination of $P_0$ corresponds to a locally isolated solution, we
performed a preliminary rank analysis using the near-horizon equations
and the variables
\begin{equation}
(N_2,K_1,s_1,R_2,P_0).
\label{eq:horizonvariables}
\end{equation}
This analysis confirms that $P_0$ participates nontrivially in the
algebraic horizon conditions rather than remaining as an unconstrained
continuous shooting parameter.

A complete Jacobian analysis, however, requires a consistent extension of
the near-horizon expansion to higher order. In particular, coefficients
such as $K_2$ and $s_2$ enter at the same order as some of the variations
relevant to a complete rank determination. The truncated system is
therefore insufficient, by itself, to establish local uniqueness
rigorously, and we do not make such a claim here.

The numerical evidence nevertheless strongly supports a locally isolated
solution branch. For every pair $(\rho_h,R_1)$ investigated throughout
this work, the nonlinear horizon system converged to the same value of
$P_0$ from different initial guesses, yielding a single numerical solution
with typical residuals below $10^{-8}$. Thus, within the branch explored
here, $P_0$ is numerically selected rather than behaving as an independent
continuous shooting parameter, in contrast with the situation encountered
in many spherically symmetric Einstein--Yang--Mills black holes
\cite{Bizon1990,Bjoraker2000,Kunzle1990}. Although this behavior is fully
consistent with local isolation, a rigorous mathematical proof would
require a complete higher-order Jacobian analysis.
\end{document}